# Roles of non-local electron-phonon coupling on the electrical conductivity and Seebeck coefficient: a TD-DMRG study

Yufei Ge [1], Weitang Li [2], Jiajun Ren [3], Zhigang Shuai [4, a]

[1] MOE Key Laboratory on Organic OptoElectronics and Molecular Engineering, Department of Chemistry, Tsinghua University, 100084 Beijing, China

[2] Tencent Quantum Lab, Tencent, Shenzhen, 518057 Guangdong, China

[3] MOE Key Laboratory of Theoretical and Computational Photochemistry, College of Chemistry, Beijing Normal University, 100875 Beijing, P. R. China

[4] School of Science and Engineering, The Chinese University of Hong Kong, Shenzhen, 518172 Guangdong, China

[a] email: shuaizhigang@cuhk.edu.cn

**Abstract:** Organic molecular materials are potential high-performance thermoelectric materials. Theoretical understanding of thermoelectric conversion in organic materials is essential for rational molecular design for efficient energy conversion materials. In organic materials, non-local electron-phonon coupling plays a vital role in charge transport and leads to complex transport mechanisms, including hopping, phonon-assist current, band, and transient localization. In this work, based on time-dependent density matrix renormalization group (TD-DMRG) method, we look at the role of non-local electron-phonon coupling on the thermoelectric conversion in organic systems described by Holstein-Peierls model. We calculate the current-current correlation and the heat current-current correlation functions. We find that (i) non-local electron-phonon coupling has very weak influence on Seebeck coefficient because of the cancellation between heat current-current correlation function and the current-current correlation function, but it has strong influence on conductivity through dynamic disorders; (ii) doping concentration has a strong influence on both conductivity and Seebeck coefficient significantly, and the optimal doping ratio to reach the highest power factor is 3%-10% fillings.

# 1 Introduction

Organic materials hold immense potential for next-generation thermoelectric devices[1,2,3], prompting significant research efforts towards designing high-performance organic thermoelectric materials over the past decades[2,4,5]. The figure of merit ($ZT$) of organic materials at room temperature has been improved by over one order of magnitude in the past decades[6]. $ZT$ value, which measures the performance of thermoelectric materials, is defined as $ZT = \frac{\alpha^2 \sigma T}{\kappa_e + \kappa_L}$, where $\alpha$, $\sigma$, $T$ and $\kappa_e$ ($\kappa_L$) represent Seebeck coefficient, electrical conductivity, temperature, and electrical (lattice) thermal conductivity. The rapid improvement in $ZT$ can be attributed to the ingenious concept named 'electron crystal, phonon glass'[3], which emphasizes the independent control of lattice thermal conductivity and electrical conductivity. Along this line, numerous techniques have been proposed[7]. However, $ZT$ values of organic materials are still low that limits application. One restriction is that conductivity and Seebeck coefficient usually behave oppositely, which limits the enhancement of power factor ($PF = \alpha^2 \sigma$) and $ZT$ value.

Essentially, Seebeck coefficient represents the 'transport entropy', which is the entropy carried by unit charge carrier in charge transport [8,9]. Therefore, to enhance thermoelectric performance further, it's necessary to interpret charge carries' thermoelectric transport process comprehensively. Obviously, electron-phonon coupling plays a vital role in thermoelectric transport. Generally speaking, electron-phonon coupling can be attributed to two types: local electron-phonon coupling, originated from intra-molecular vibration and non-local electron-phonon coupling comes from inter-molecular vibration. Previous studies have systematically revealed the effect of local electron-phonon coupling which decreases conductivity and enhances Seebeck coefficient by bandwidth narrowing effect[10–12]. Meanwhile, non-local electron-phonon coupling which can lead to the formation of solitons, polarons and bipolarons[13,14], influencing charge transport[15,16], exciton dynamic[17] and spectral properties[18] significantly, hasn't been fully[19] studied in thermoelectric transport.

According to previous research, non-local electron-phonon coupling introduces non-local dynamic disorder and involves various transport regime, including (i) hopping regime[5,20] where charge carries are fully localized at one molecule because of strong local electron-phonon coupling, and 'hop' from one molecule to another, (ii) phonon-assist current regime[21,22] where exists large thermal fluctuation of transfer integral induced by non-local electron-phonon coupling, which enhances mobility, (iii) band regime[23,24] where charge carriers move 'wave-like' and are scattered by non-local phonons and impurities. (iv) transient localize regime[25] where charge carriers are time-dependently localized at several molecules by the motion of molecules (dynamic disorder), and (v) intermediate regime[19] which belongs to none of the regimes mentioned above.

The involved complex transport regimes hinder the comprehensive understanding of non-local electron-phonon coupling on thermoelectric transport for a long time because of lacking numerical exact simulation method[15]. Fortunately, the time-dependent density matrix renormalization group[26–28] (TD-DMRG) has the potential to overcome the difficulty in simulation[19,29], and in our previous work the computational methods for calculating conductivity and Seebeck coefficient by TD-DMRG have been developed, which is both valid and numerically exact[12].

In this work, we will adopt TD-DMRG to investigate thermoelectric transport in organic materials, and reveal the influence of non-local electron-phonon coupling in different charge transport regimes.

## 2 Model and Computational Approach

Holstein-Peierls model[13,30] captures the effect of local and non-local electron-phonon coupling in organic materials[31]. In this work, we consider $N$ molecules in a one-dimensional chain with lattice constant $\Omega$ and each molecule contributes one molecular orbital (LUMO or HOMO corresponding to n-type or p-type doping) to thermoelectric transport process. With periodic boundary condition, the Hamiltonian contains the following five parts:

$$\hat{H} = \hat{H}_e + \hat{H}_{e-ph1} + \hat{H}_{e-ph2} + \hat{H}_{ph1} + \hat{H}_{ph2} \quad (1)$$

Here, the electron part reads:

$$\hat{H}_e = \sum_j \epsilon_j \hat{a}_j^\dagger \hat{a}_j + \sum_j \tau_{j,j+1}(\hat{a}_{j+1}^\dagger \hat{a}_j + \hat{a}_j^\dagger \hat{a}_{j+1}) \quad (2)$$

In the electron part, $\hat{a}_j^\dagger$ and $\hat{a}_j$ are creation and annihilation operator of the $j$th molecule's orbital. Considering the tradition of DMRG, we denote it as 'site $j$' below for brevity. $\epsilon_j$ and $\tau_{j,j+1}$ are orbital energy at site $j$ and transfer integral between site $j$ and $j+1$, respectively. For simplicity, we set $\epsilon_j = 0$ and $\tau_{j,j+1} = \tau$.

Holstein model captures the effect of intra-molecular electron-vibration coupling (i.e. local coupling), including bandwidth narrowing effect and reorganization energy[11,32]. $\hat{H}_{e-ph1}$ and $\hat{H}_{ph1}$ correspond to the Holstein model. Here, local electron-phonon coupling term is

$$\hat{H}_{e-ph1} = \sum_{j,n} \hbar g_{H,n} \omega_{H,n}(\hat{b}_{jn}^\dagger + \hat{b}_{jn})\hat{a}_j^\dagger \hat{a}_j \quad (3)$$

and local phonon energy term is

$$\hat{H}_{ph1} = \sum_{j,n} \hbar \omega_{H,n}\left(\hat{b}_{jn}^\dagger \hat{b}_{jn} + \frac{1}{2}\right) \quad (4)$$

$\hat{b}_{jn}^\dagger$ and $\hat{b}_{jn}$ are creation and annihilation operator of the phonons corresponding to the $n$th vibration mode at site $j$. $\omega_{H,n}$ and $g_{H,n}$ are the phonon frequency and the electron-phonon coupling constant of the $n$th intra-molecular vibration mode. The reorganization energy is defined as $\lambda = \sum_n g_{H,n}^2 \omega_{H,n}$, which represents the strength of local electron-phonon coupling. For simplicity, we use four modes, $\omega_H = $ 40meV, 120meV, 200meV, 280meV and $g_H = 1.247, 0.645, 0.2311, 0.0792$.

Peierls model captures the effect of inter-molecular electron-vibration coupling (i.e. non-local coupling), leading to thermal fluctuation of transfer integral[15]. $\hat{H}_{e-ph2}$ and $\hat{H}_{ph2}$ correspond to Peierls model. Here, non-local electron-phonon coupling term is

$$\widehat{H}_{e-ph2} = \sum_j \hbar g_P \omega_P (\hat{c}_j^\dagger + \hat{c}_j)(\hat{a}_j^\dagger \hat{a}_{j+1} + \hat{a}_{j+1}^\dagger \hat{a}_j) \tag{5}$$

And non-local phonon energy is

$$\widehat{H}_{ph2} = \sum_j \hbar \omega_P \left( \hat{c}_j^\dagger \hat{c}_j + \frac{1}{2} \right) \tag{6}$$

Here, $\hat{c}_j^\dagger$ and $\hat{c}_j$ are creation and annihilation operator of the phonons corresponding to the inter-molecular vibration between site $j$ and site $j+1$. $\omega_P$ and $g_P$ are vibration frequency and coupling constant of inter-molecular vibration mode. For simplicity, one inter-molecular vibration mode $\omega_P = 10 meV$ is adopted[31,33]. Thermal fluctuation of transfer integral $\Delta V$ reflects the strength of non-local electron-phonon coupling and dynamic disorder, and the relationship between $g_P$ and $\Delta V$ is

$$\Delta V = g_P \omega_P \sqrt{\coth \frac{\omega_P}{2k_B T}} \tag{7}$$

The conductivity $\sigma$ and Seebeck coefficient $\alpha$ are calculated via Kubo formula[10,11]:

$$\begin{cases} \sigma = \dfrac{1}{k_B T V} \int_0^{+\infty} \text{Re } C_1(t) dt \\ \alpha = \dfrac{1}{T} \dfrac{\int_0^{+\infty} \text{Re } C_2(t) dt}{\int_0^{+\infty} \text{Re } C_1(t) dt} \end{cases} \tag{8}$$

Where $V$ is the volume of the unit cell, $T$ is temperature, $k_B$ is Boltzmann constant, and the current-current correlation function $C_1(t)$ and heat current - current correlation function $C_2(t)$ are defined as bellow[34]:

$$\begin{cases} C_1(t) = \text{Tr}[\hat{\rho}_0 e^{i\widehat{H}t/\hbar} \hat{J}_e e^{-i\widehat{H}t/\hbar} \hat{J}_e] \\ C_2(t) = \text{Tr}[\hat{\rho}_0 e^{i\widehat{H}t/\hbar} \hat{J}_Q e^{-i\widehat{H}t/\hbar} \hat{J}_e] \end{cases} \tag{9}$$

Here, $\hat{J}_e$ represents electrical current operator:

$$\hat{J}_e = -\frac{i}{\hbar} e\Omega \sum_j \widehat{T}_{j,j+1}(\hat{a}_{j+1}^\dagger \hat{a}_j - \hat{a}_j^\dagger \hat{a}_{j+1}) \tag{10}$$

$\hat{J}_Q$ represents heat current operator:

$$\hat{J}_Q = \hat{J}_Q^I + \hat{J}_Q^{II} + \hat{J}_Q^{III} \tag{11}$$

$$\hat{J}_Q^I = -\frac{i}{\hbar}\Omega \sum_j \hat{T}_{j,j+1}\hat{T}_{j+1,j+2}\bigl(\hat{a}_{j+2}^\dagger \hat{a}_j - \hat{a}_j^\dagger \hat{a}_{j+2}\bigr) \tag{12}$$

$$\hat{J}_Q^{II} = -\frac{i}{\hbar}\Omega \sum_j \left[\frac{1}{2}\bigl(\hat{E}_j + \hat{E}_{j+1}\bigr) - \mu\right]\hat{T}_{j,j+1}\bigl(\hat{a}_{j+1}^\dagger \hat{a}_j - \hat{a}_j^\dagger \hat{a}_{j+1}\bigr) \tag{13}$$

$$\hat{J}_Q^{III} = \frac{i}{2\hbar}\Omega \sum_j \hbar^2 g_P \omega_P^2 \bigl(\hat{c}_j^\dagger - \hat{c}_j\bigr)\bigl(\hat{a}_{j+1}^\dagger \hat{a}_j + \hat{a}_j^\dagger \hat{a}_{j+1}\bigr) \tag{14}$$

Here, $\hat{J}_Q^I$, $\hat{J}_Q^{II}$ and $\hat{J}_Q^{III}$ represent heat current originated from electron's kinetic energy, electron's potential energy and inter-molecular vibration. And $\hat{E}_j$ is the on-site energy operator corrected by local phonon and $\hat{T}_{j,j+1}$ is the transfer integral operator corrected by non-local phonon:

$$\hat{E}_j = \epsilon_j + \sum_n \hbar g_{H,n} \omega_{H,n}\bigl(\hat{b}_{jn}^\dagger + \hat{b}_{jn}\bigr) \tag{15}$$

$$\hat{T}_{j,j+1} = \tau_{j,j+1} + \hbar g_P \omega_P \bigl(\hat{c}_j^\dagger + \hat{c}_j\bigr) \tag{16}$$

Grand canonical ensemble is adopted and density operator is

$$\hat{\rho}_0 = \frac{1}{Z} e^{-\beta(\hat{H} - \mu \hat{N}_e)} \tag{17}$$

Here, partition function is $Z = \mathrm{Tr}\bigl[e^{-\beta(\hat{H}-\mu\hat{N}_e)}\bigr]$ and electron number operator is $\hat{N}_e = \sum_j \hat{a}_j^\dagger \hat{a}_j$. Doping ratio is defined as $c = \mathrm{Tr}[\hat{\rho}_0 \hat{N}_e]/N$. Unless otherwise specified, the lattice constant $\Omega = 10\ a.u.$, the doping ratio is set as $c = 0.001$, and the temperature is set as $T = 300K$.

The numerical calculation is carried out by transform Eq. (9) into

$$\begin{cases} C_1(t) = \langle \Psi_\beta | e^{i\hat{H}t/\hbar} \hat{J}_e e^{-i\hat{H}t/\hbar} \hat{J}_e | \Psi_\beta \rangle \\ C_2(t) = \langle \Psi_\beta | e^{i\hat{H}t/\hbar} \hat{J}_Q e^{-i\hat{H}t/\hbar} \hat{J}_e | \Psi_\beta \rangle \end{cases} \tag{18}$$

Here, $|\Psi_\beta\rangle$ is thermal state obtained by purification method [27]. As presented in Fig. 1, the electron sites and phonon sites of molecular are mapped to a one-dimension chain. The green circles, purple circles, and blue circles represent tensors of thermal states $|\Psi_\beta\rangle$ and $\langle\Psi_\beta|$, corresponding to electron sites, non-local phonon sites, and local phonon sites. Note that electron sites are represented by spin basis after Jordan-Wigner transform. Then, matrix product operator (MPO) $\hat{J}_e$ is applied to $|\Psi_\beta\rangle$. The orange diamonds represent time evolution of $\hat{J}_e|\Psi_\beta\rangle$ and $\langle\Psi_\beta|$, where time-dependent variational principle with a projector-splitting algorithm (TDVP-PS) is adopted [28]. Finally, correlation functions $C_1(t)$ and $C_2(t)$ are obtained by calculating expectation value of $\hat{J}_e$ and $\hat{J}_Q$ (yellow squares). The sequence of algorithm is indicated by blue arrows in Fig. 1. The details of parameters' selections are presented in Appendix A.

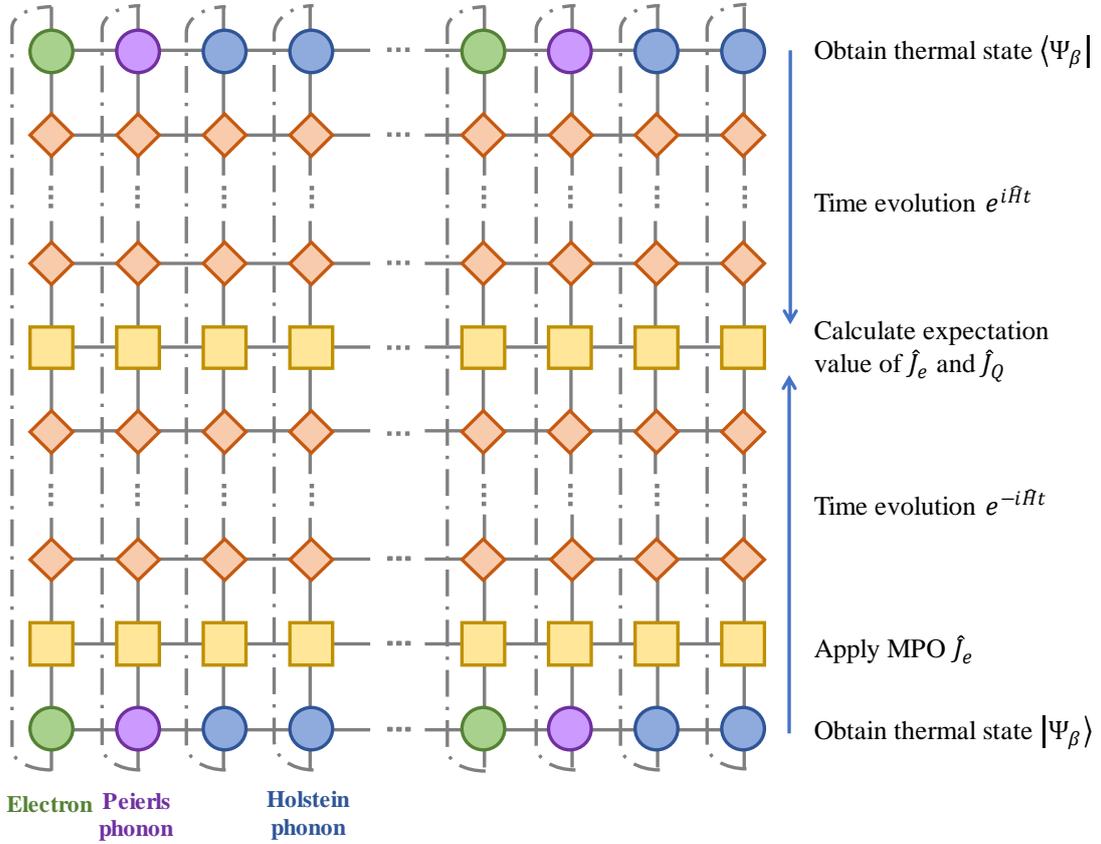

Fig. 1: Schematic diagram of numerical calculation progress.

## 3 Results and Discussions

As mentioned above, the effect of non-local electron-phonon coupling should be studied across five different regimes[19]. The five regimes are hopping regime, phonon-assist current regime, band regime, transient localize regime, and intermediate regime, as presented in Fig. 2.

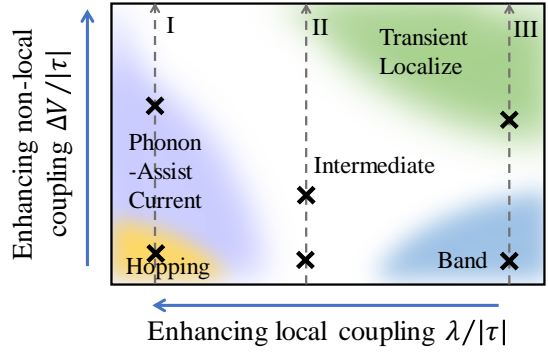

Fig. 2 Five transport regimes in organic materials. Yellow part is hopping regime, where $|\tau| \ll \lambda$ and $\Delta V \ll |\tau|$; purple part is phonon-assist current regime, where $|\tau| \ll \lambda$ and $\Delta V$ is comparable to or larger than $|\tau|$; blue part is band regime, where $|\tau| \gg \lambda$ and $\Delta V \ll |\tau|$; green part is transient localize regime, where $|\tau| \gg \lambda$ and $\Delta V$ is comparable to or larger than $|\tau|$; white part is intermediate regime, where $|\tau| \sim \lambda$ and $\Delta V \sim |\tau|$. The gray dashed lines (I), (II) and (III) corresponding to the parameter selection in Fig. 3. The black crosses corresponding to the representative parameters of five transport regimes, which are adopted in Fig. 4-6.

First, we study the influence of non-local electron-phonon coupling under three representative cases, namely, strong, intermediate and weak local electron-phonon coupling. Figures 3(a)-3(c) correspond to the case of strong local electron-phonon coupling ($|\tau| \ll \lambda$). When $\Delta V$ is small, the conductivity $\sigma$ increases monotonically with increasing $\Delta V$, matching well with the analytical solution in hopping limit (Appendix B). The Seebeck coefficient $\alpha$ follows the formula for hopping limit[35]: $\alpha = -\frac{k_B}{e}\ln\left(\frac{1-c}{c}\right)$, remaining independent of $\Delta V$. Additionally, the mean free path $l_{\mathrm{mfp}}$ is much smaller than the lattice constant, suggesting a hopping transport mechanism. Here, mean free path $l_{\mathrm{mfp}}$ is defined as[36]:

$$l_{\mathrm{mfp}} = \left[\frac{1}{e^2 n_e}C_1(0)\right]^{\frac{1}{2}} \int_0^{+\infty} dt \left|\frac{\mathrm{Re}\, C_1(t)}{\mathrm{Re}\, C_1(0)}\right| \qquad (19)$$

When $\Delta V$ increases further, $\sigma$ exhibits a steady rise while moving away from the hopping behavior. Simultaneously, $\alpha$ experience a decrease due to the widening of the bandwidth, which shall be illustrated in Fig. 4(a) and 4(b). Considering $l_{\text{mfp}} \sim \Omega$, the dominant charge transport mechanism is phonon-assisted current when $\Delta V$ attains large values.

In Fig. 3(g)-3(i), we examine the case of weak local electron-phonon coupling where $\lambda \ll |\tau|$. For small values of $\Delta V$, the conductivity experiences a rapid decrease as $\Delta V$ increases, matching well with the predictions of band theory[37]. The Seebeck coefficient matches the constant behavior predicted by band theory (Appendix C) when $\Delta V = 0$ and exhibits a slight increase. The observation $l_{\text{mfp}} \gg \Omega$ suggests a band-like behavior as well. As $\Delta V$ increases further, the conductivity continues to decrease, while the Seebeck coefficient starts to decrease due to the broadening of the bandwidth (see Fig. 4(e) and 4(f)). Considering $l_{\text{mfp}} \sim \Omega$, the transport mechanism conforms to the characteristics of transient localization[38].

In Fig. 3(d)-3(f), we investigate the intermediate case where the transfer integral $|\tau|$ is comparable to the reorganization energy $\lambda$. As the parameter $\Delta V$ increases, the conductivity exhibits a consistent and monotonic decrease. The Seebeck coefficient, on the other hand, remains constant initially and then shows a slight increase with increasing $\Delta V$. Additionally, $l_{\text{mfp}}$ consistently remains comparable to the lattice constant $\Omega$ throughout the analysis. These behaviors are different from the mechanism mentioned above and are attributed to intermediate transport regime[19].

Moreover, it should be noted that, under different strength of local electron-phonon coupling, the change of $\Delta V$ influences the value of conductivity significantly (more than ten times) while changes the absolute value of Seebeck coefficient little (less than 10%).

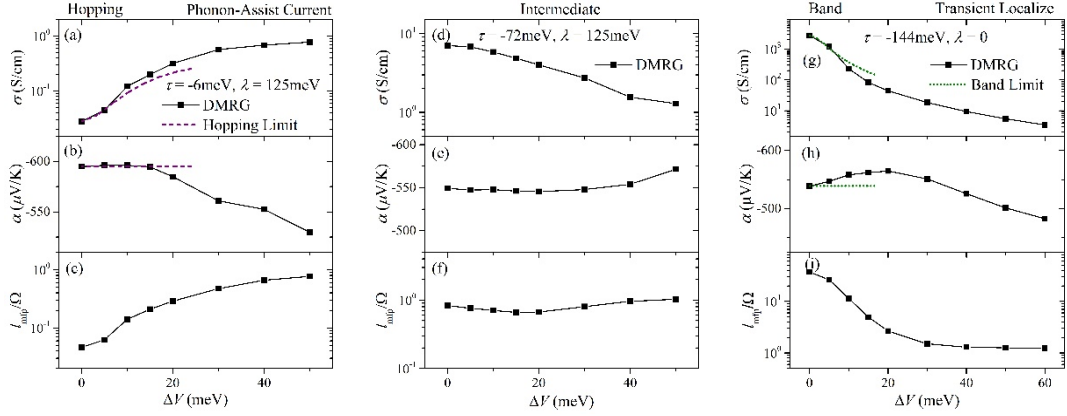

Fig. 3 Influence of non-local electron-phonon coupling $\Delta V$ on conductivity $\sigma$, Seebeck coefficient $\alpha$ and mean free path $l_{\text{mfp}}$ under different local electron-phonon coupling. (a)-(c), (d)-(f), and (g)-(i) corresponding to strong, intermediate, and weak local electron-phonon coupling, respectively. Here, $c = 0.001$ and $T = 300K$.

Based on transport regimes identified in Fig. 3, we select the following representative parameters to cover the five transport regimes mentioned above, specifically, $\tau = -6meV$, $\lambda = 125meV, g_P = 0.25$ for hopping regime, $\tau = -6meV$, $\lambda = 125meV, g_P = 1.5$ for phonon-assist current regime, $\tau = -72meV$, $\lambda = 125meV, g_P = 0.25$ and $\tau = -72meV$, $\lambda = 125meV, g_P = 1.0$ for intermediate regime (denoted as intermediate regime 1 and 2), $\tau = -144meV$, $\lambda = 0, g_P = 0.25$ for band regime, and $\tau = -144meV$, $\lambda = 0, g_P = 2.5$ for transient localize regime. The parameters correspond to the black crosses in Fig. 2.

The influence of non-local electron-phonon coupling on one-particle spectral density function $A(k, E)$ across five transport regimes are plotted in Fig. 4. Here[15],

$$A(k, E) = \frac{1}{N\pi} \sum_{jl}^{N} e^{ik\Omega(j-l)} \int_0^{+\infty} \text{Tr}[\hat{a}_j(t)\hat{a}_l^\dagger(0)] e^{iEt} dt \qquad (20)$$

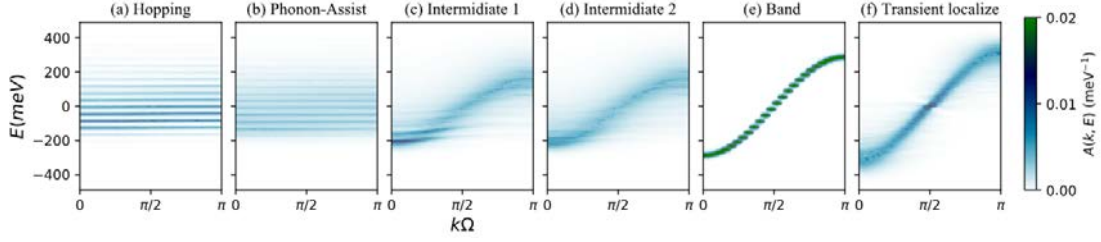

Fig. 4 One-particle spectral density function when $T = 300K$ in (a) hopping regime, (b) phonon-assist current regime, (c) intermediate regime 1, (d) intermediate regime 2, (e) band regime, and (f) transient localize regime.

In Fig. 4(a), because of the bandwidth narrowing effect[39] introduced by local coupling, $A(k, E)$ is so narrow and becomes discrete in hopping regime, indicating the localization of charge carriers. When non-local coupling $\Delta V$ grows, the transport behavior comes to phonon-assist current regime. As presented in Fig. 4(b), $A(k, E)$ becomes more dispersive[22]. In intermediate regime 1 and 2 presented in Fig. 4(c) and 4(d), $A(k, E)$ is much wider and more dispersive than Fig. 4(a) and 4(b) because of larger transfer integral. In band regime presented in Fig. 4(e), $A(k, E)$ is wide and corresponds to the formula of energy band, i.e. $E(k) = 2\tau \cos k\Omega$. As non-local coupling increases, the transport mechanism comes to transient localize regime[19,40], which is shown in Fig. 4(f). Here, $A(k, E)$ is more dispersive than $A(k, E)$ in Fig. 4(e).

The difference between Fig. 4(a) and 4(b), Fig. 4(c) and 4(d), Fig. 4(e) and 4(f), show that increasing non-local electron-phonon coupling ($\Delta V$) just modifies the shapes of $A(k, E)$ slightly by making $A(k, E)$ more dispersive. While the change of local coupling influences $A(k, E)$ obviously, as presented in Fig. 4(a), 4(c) and 4(e). These behaviors are essential as the density of state (DOS) can be calculated as below[15]:

$$D(E) = \frac{1}{V} \sum_k A(k, E) \qquad (21)$$

According to general formula for Seebeck coefficient:

$$\alpha = \int_{-\infty}^{+\infty} dE \frac{(E - \mu)}{eT} \frac{\sigma(E)}{\sigma} \left( -\frac{\partial f}{\partial E} \right) \qquad (22)$$

where

$$\sigma = \int_{-\infty}^{+\infty} dE \sigma(E) \left( -\frac{\partial f}{\partial E} \right) \tag{23}$$

and

$$\sigma(E) = e\mu_c(E)D(E)k_B T \tag{24}$$

Here, $f = \frac{1}{1+\exp\left(\frac{E-\mu}{k_B T}\right)}$ is Fermi-Dirac distribution function and $\mu_c(E)$ is carrier mobility. According to previous research, when DOS is narrow, Seebeck coefficient is a constant $\alpha = -\frac{k_B}{e}\ln\left(\frac{1-c}{c}\right)$ independent of $\Delta V$, as shown in hopping regime presented in Fig. 3(b). When DOS is wide, the behavior of Seebeck coefficient can be qualitatively interpreted via Mott's formula[41], which indicates that the value Seebeck coefficient is decided by the shape of DOS, and wider DOS leads to lower Seebeck coefficient:

$$\alpha \approx \frac{\pi^2}{3}\frac{k_B}{e}k_B T \left.\frac{d \ln D(E)}{dE}\right|_{E=\mu} \tag{25}$$

Therefore, (i) the fact that changing non-local coupling ($\Delta V$) just modifies $A(k,E)$ slightly explains why the influence of $\Delta V$ only has small influence on Seebeck coefficient in Fig. 3. (ii) the DOS are slightly broadened as $\Delta V$ increases when $|\tau| \ll \lambda$ and $|\tau| \gg \lambda$ but remains unchanged as $\Delta V$ increases when $|\tau| \sim \lambda$. These explain the decrease of Seebeck coefficient as $\Delta V$ grows in phonon-assist current regime (Fig.3(b)) and transient localize regime (Fig. 3(h)), and the constant behavior of $\alpha$ in intermediate regime (Fig. 3(e)).

The temperature dependence of conductivity and Seebeck coefficient in various transport regimes are depicted in Fig. 5. Specifically, in hopping regime shown in Fig. 5(a) and 5(b), we observe a decrease in conductivity with increasing temperature due to the enhancement of bandwidth narrowing effect. Meanwhile, the Seebeck coefficient remains constant as temperature increases, attributed to the narrow polaron band[8,11]. This behavior aligns with the formula $\alpha = -\frac{k_B}{e}\ln\left(\frac{1-c}{c}\right)$, which is temperature-independent. In phonon-assisted current regime, the conductivity also decreases as temperature rises, while the Seebeck coefficient remains relatively unchanged. In intermediate transport regime shown in Fig. 5(c) and 5(d), the conductivity experiences a rapid decrease with increasing temperature. Simultaneously, the Seebeck coefficient is roughly independent of temperature. In the transient localize regime and band regime shown in Fig. 5(e) and 5(f), the conductivity decreases with increasing temperature due to the growing dynamic disorder. For Seebeck coefficient, we observe a roughly linear increase with temperature, which can be interpreted through Mott's Formula[41] (Eq. (25)).

Note that the strength of non-local electron-phonon coupling $\Delta V$ increases as temperature increases (Eq. (7)). Therefore, the temperature dependence also suggests non-local coupling has significant influence on electrical conductivity but little effect on Seebeck coefficient.

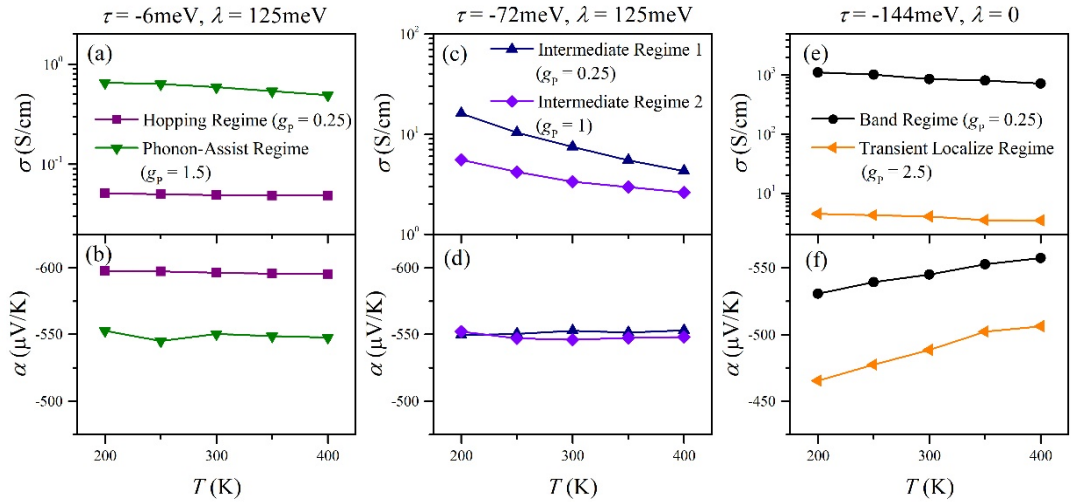

Fig. 5. Temperature dependence of conductivity $\sigma$ and Seebeck coefficient $\alpha$ in different transport regimes. Here, $c = 0.001$. Note that in (a) and (b), $\tau = -6meV$ and $\lambda = 125meV$; in (c) and (d), $\tau = -72meV$ and $\lambda = 125meV$; in (e) and (f), $\tau = -144meV$ and $\lambda = 0$.

Fig. 6 illustrates the dependence of transport coefficients on doping ratio $c$ in various transport regimes. Take electron doping (Fig. 6(a)-6(c)) as an example, we observe that the conductivity consistently increases as the doping ratio increases, while the Seebeck coefficient exhibits an opposite trend. Throughout all transport regimes, the conductivity follows a proportional relationship $\sigma \propto \ln c$, and Seebeck coefficient roughly follows $\alpha = A \ln c + B$, where $A$ and $B$ are constants. Additionally, the absolute values of conductivity vary significantly across different transport regimes.

Typically, we expect higher conductivity in the band regime, intermediate regime, and transient localize regime due to large effective transfer integrals, which are reflected in the spectral density functions shown in Fig. 4(c)-4(f). Different from the huge difference in the value of conductivities in different transport regimes, the values of Seebeck coefficient are similar in different transport regimes under a fixed doping ratio. Notably, the doping ratio $c$ influences both conductivity and Seebeck coefficient significantly, which can be interpreted in Eq. (22) and (23), where doping ratio (related to chemical potential $\mu$) are directly involved in the expression of $\sigma$ and $\alpha$. The power factors, $PF = \alpha^2 \sigma$, are plotted in Fig. 6(c). The optimal electron doping ratio for achieving the highest power factor is approximately 3%-10% across all transport regimes. Moreover, the case of hole doping is also investigated in Fig. 6(d)-6(f), demonstrating impressive similarities to the electron doping case as expected.

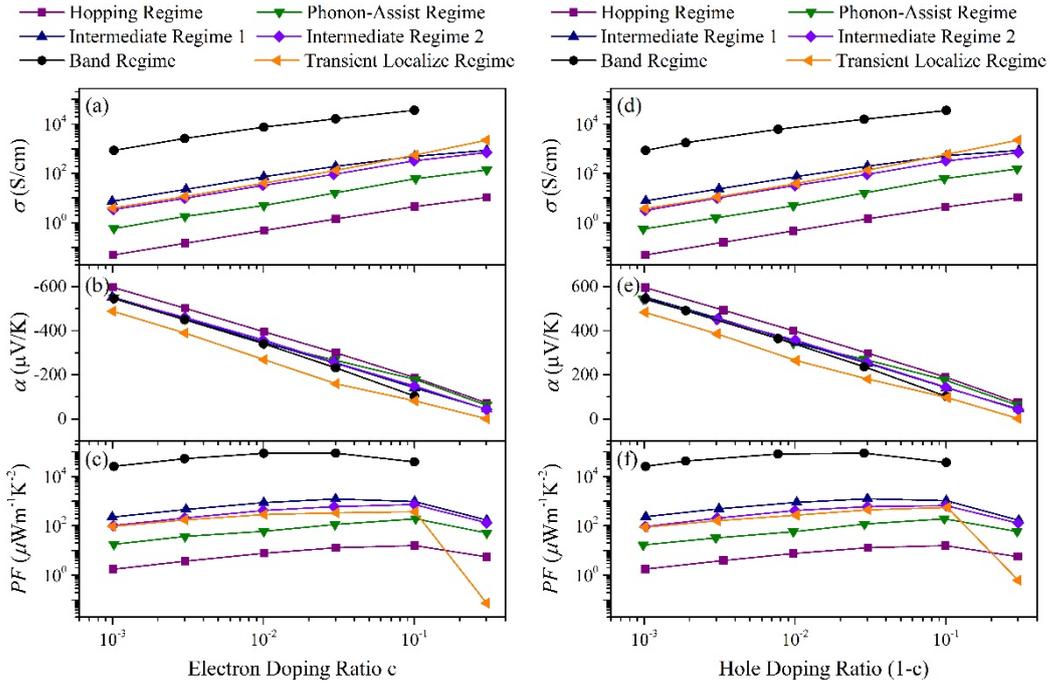

Fig. 6 Dependence of conductivity $\sigma$, Seebeck coefficient $\alpha$ and power factor $PF$ on doping ratio $c$ when $T = 300K$. (a)-(c) corresponding to electron doping and (d)-(f) corresponding to hole doping.

## 4 Conclusions

In summary, we conducted a comprehensive investigation of non-local electron-phonon coupling's influence on thermoelectric transport in organic materials, adopting TD-DMRG method that overcomes the limitations of previous approaches. Notably, we find that non-local coupling influences conductivity significantly while has little effect on Seebeck coefficient, which can be interpreted via the change of DOS and the general expression in thermoelectric transport. Meanwhile, doping ratio influence both conductivity and Seebeck coefficient significantly, and the optimal doping ratio for highest power factor is making HOMOs (LUMOs) 3%-10% filled by holes(electrons).

Therefore, our work indicates an experimental strategy valid in all transport regimes for higher thermoelectric power factor: we can enhance conductivity through rational molecular design to control transfer integral or the impact of electron-phonon coupling for higher mobility and then achieve an optimal doping ratio that balance conductivity and Seebeck coefficient for highest power factor.

## Appendix A: Details of numerical calculations

We carry out calculations via python package RENORMALIZER. The computational parameters adopted are listed in Table A1.

Table A1 parameters adopted in numerical calculation

| Parameters | Value selection |
| --- | --- |
| Virtual bond dimension $D$ | 64 |
| Size of local phonon basis $d_H$ | 9 |
| Size of non-local phonon basis $d_P$ | 60 |
| Site number | 16 |
| Imaginary time evolution steps | 300 |
| Time step $\Delta t$ of real time evolution | ~25 a.u. |
| Damping function for band regime | $\gamma(t) = \exp\left[-\left(\frac{t^2}{t_0^2}\right)\right], t_0 = \frac{\hbar}{\gamma_0},$ $\gamma_0 = 2meV$ |

The virtual bond dimension $D$ and the size of non-local phonon basis $d_P$ are tested as well, which is plotted in Fig. A1.

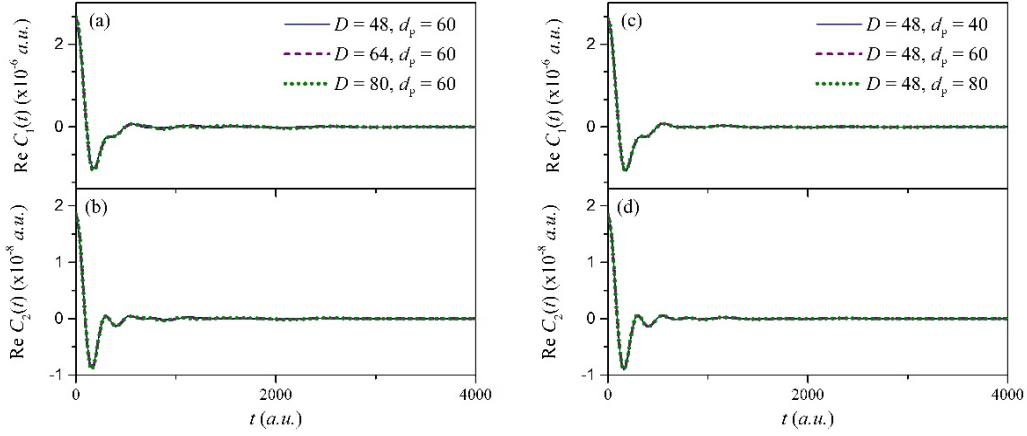

Fig A1 (a), (b) convergence of virtual bond dimension $D$, (c), (d) convergence of the size of non-local phonon basis $d_P$ in our work.

## Appendix B: Hopping behavior

The conductivity in hopping limit is calculated via:

$$\sigma = n_e e \mu_e \qquad (B1)$$

Where electron mobility $\mu_e$ is calculated via[19]:

$$\mu_e = \frac{e\Omega^2}{k_B T} \int_{-\infty}^{\infty} [\tau^2 + (g_P \omega_P)^2 f(\omega_P, t)] e^{-\Gamma(t)} dt \qquad (B2)$$

$$\Gamma(t) = 2\sum_n g_{H,n}^2 [1 + 2N(\omega_{H,n}) - f(\omega_{H,n}, t)] + 4g_P^2 [1 + 2N(\omega_P) - f(\omega_P, t)] \quad (B3)$$

$$f(\omega, t) = (1 + N(\omega))e^{-i\omega t} + N(\omega)e^{i\omega t} \qquad (B4)$$

$$N(\omega) = \frac{1}{e^{\omega/k_B T} - 1} \qquad (B5)$$

Seebeck coefficient in hopping limit is calculated via[35]:

$$\alpha = -\frac{k_B}{e} \ln\left(\frac{1-c}{c}\right) \qquad (B6)$$

## Appendix C: Band Limit

The band limit in the manuscript is calculated via Boltzmann Transport Equation[37]:

$$\sigma = e^2 \sum_k \left(-\frac{\partial f}{\partial E_k}\right) v_k v_k \theta_k \qquad (C1)$$

$$\alpha = \frac{e}{\sigma} \sum_k \frac{E_k - \mu}{T} \left(-\frac{\partial f}{\partial E_k}\right) v_k v_k \theta_k \tag{C2}$$

Where $k$ is the wave vector, $\theta_k$ is the relaxation time, group velocity $v_k = \frac{1}{\hbar} \frac{\partial E_k}{\partial k}$ and $f$ is the Fermi-Dirac distribution function. In rigid band approximation, the energy band of electrons is:

$$E_k = 2\tau \cos k\Omega \tag{C3}$$

The total relaxation time can be expressed as:

$$\frac{1}{\theta_{total}(k)} = \frac{1}{\theta_{imp}(k)} + \frac{1}{\theta_{ph}(k)} \tag{C4}$$

$\theta_{imp}(k)$ is set as a constant $\theta_0$ which satisfies:

$$\sigma(g_P = 0, \lambda = 0) = e^2 \sum_k \left(-\frac{\partial f}{\partial E_k}\right) v_k v_k \theta_0 \tag{C5}$$

According to deformation potential approximation[24]:

$$\frac{1}{\theta_{ph}(k)} = \sum_{k'} \frac{2\pi}{\hbar} \frac{k_B T E_1^2}{C_{ii}} \delta(E_{k'} - E_k)(1 - \cos \gamma) \tag{C6}$$

Here, $C_{ii}$ is elastic constant, $E_1$ is deformation potential constant, and $\gamma$ is scattering angle. For one-dimensional system with $E - k$ relationship presented in Eq. (C3), $\gamma = 0$ or $\pi$ when $\delta(E_{k'} - E_k)$ is nonzero. Considering the definition of $E_1$ and $C_{ii}$, $\frac{1}{\theta_{ph}(k)}$ is a constant independent of $k$ and proportional to $g_P$, i.e. $\frac{1}{\theta_{ph}(k)} = A_g g_P$, where the factor $A_g$ can be calculated by second data point presented in Fig. 2(g).

## Acknowledgement:

This work is supported by Shenzhen Science and Technology Program and by Guangdong Provincial Basic and Applied Research Grant. Jiajun Ren is also supported by the NSFC via Grant Number 22273005.